# Colloidal superlattices for unnaturally high-index metamaterials at broadband optical frequencies

*Seungwoo Lee*[1*]

[1]SKKU Advanced Institute of Nanotechnology (SAINT) & School of Chemical Engineering, Sungkyunkwan University (SKKU), Suwon 440-746, Republic of Korea

[*]Email: seungwoo@skku.edu

**Abstract**

The recent advance in the assembly of metallic nanoparticles (NPs) has enabled sophisticated engineering of unprecedented light-matter interaction at the optical domain. In this work, I expand the design flexibility of NP optical metamaterial to push the upper limit of accessible refractive index to the unnaturally high regime. The precise control over the geometrical parameters of NP superlattice monolayer conferred the dramatic increase in electric resonance and related effective permittivity far beyond the naturally accessible regime. Simultaneously, effective permeability, another key factor to achieving high refractive index, was effectively suppressed by reducing the thickness of NPs. By establishing this design rule, I have achieved unnaturally high refractive index (15.7 at the electric resonance and 7.3 at the quasi-static limit) at broadband optical frequencies (100 THz ~ 300 THz). I also combined this NP metamaterial with graphene to electrically control the high refractive index over the broad optical frequencies.

## 1. Introduction

The recent experimental success in the assembly of metallic nanoparticles (NPs) in turn have stimulated theoretical predictions of unnatural light-matter interaction at the optical domain, under the umbrella of the available structural architectures [1−9]. Much of these developments have focused on (i) the proximity coupling effect between individual NPs on the bulk properties [1,4], (ii) the effect of NP assembly's geometry (periodic layer or clusters) on magnetic resonance [2,3,5,7,8], and (iii) the orientation effect of anisotropic NPs on the bulk properties [6,9]. As a result, several unnatural light-matter interactions in the optical domain, including epsilon-near-zero, negative refractive index, a magnetic mirror, and perfect absorption, have been suggested by NP assemblies [1−14].

Here, I extend such strategy to achieve unnaturally high refractive index (*n*) at broadband optical frequencies (15.7 at electric resonance frequency and 7.3 at the quasi-static regime). By optimizing the design strategy with metallic NP superlattice, I successfully maximize the polarization and the resultant effective permittivity (*ε*), while simultaneously minimizing magnetization. Thereby, the obtainable refractive index in the broadband optical domain can be effectively extended far beyond the upper limit, which is achievable with naturally occurring materials. Along this direction, the assignment of effective thickness was carefully carried out through confirming the invariant effective refractive index according to the number of NP superlattice layer. I also show the applicability of this "electric" optical metamaterial to electrically active modulation of high refractive index through the hybridization of graphene. This proposed method expands the design space of accessible optical metamaterial and thus contributes to obtaining a fruitful diversity of light-matter interaction at the optical domain.

## 2. Metallic NP superlattice as a metamaterial in the optical domain

The key of achieving high refractive index is to maximize the effective permittivity ($\varepsilon = 1 + (P/\varepsilon_0 E)$, where $P$ and $\varepsilon_0$ are polarization and permittivity in free space, respectively), while simultaneously the diamagnetic effect and the resultant reduction of effective permeability ($\mu = 1 + (M/H)$, where $M$ denotes magnetization) should be suppressed as possible [15−17]. Thus, the polarization and magnetization within the artificial medium should be maximized and suppressed respectively toward unnaturally high refractive index. Moreover, bearing the long wavelength limit in mind, the control of these polarization and magnetization at the optical regime should be performed by using sub-100-nm artificial 'atoms' (meta-atoms).

The versatile assembly of metallic NP (20 ~ 100 nm in size) into the superlattice enables the sophisticated engineering of both the polarization and magnetization at the optical domain [1−14]. Especially, the dipolar interaction between metallic NPs can affect the polarization and associated effective permittivity, whereas the diamagnetic effect can be precisely controlled by adjusting the ratio of skin depth ($\delta$) to the thickness ($d$) of NP [1,4,8]. Thereby, once the geometry of the superlattice is fixed, both polarization and magnetization can be explicitly tuned with respect to the NP motifs. Also, the gap between NPs can provide the additional degree of freedom in terms of tuning the polarization. As described below, I address the question how the superlattice of the rationally chosen metallic NPs can be indeed used to realize the "electric" optical metamaterials with unnaturally high refractive index.

## 3. Results and Discussions

### 3.1 Method

To begin with, I systematically quantified the effect of NP shape on both polarization and magnetization of artificial medium. Figure 1 summarizes the accessed pool of metallic NP motifs benefitting from the rapidly growing field of chemical synthesis [18,19]. The current work is mainly oriented toward the use of Au ($\delta$ ~ 25 nm at the optical domain) instead of silver (Ag), since much more robust to the external oxidation.

All results reported in this work were calculated by the retrieval of effective parameter (i.e., $s$-parameter extraction method based on effective medium theory) combined with the numerical simulation of scattering parameters (i.e., finite-difference time-domain (FDTD) powered by CST microwave studio) [16,17]. During numerical simulation, the complex permittivity of Au was assumed to follow Drude-critical model; that of other dielectric materials (e.g., organic ligands) was measured by ellipsometry [20]. In the case of graphene, the calculated optical conductivity by Kubo formula was converted into the effective complex permittivity by using volumetric permittivity approach [12,21]: the employed carrier relaxation time and the thickness of graphene were 250 fs and 1 nm, respectively. During the retrieval of effective parameters, the use of homogenization theory can be justified, since both the size of meta-atom and the thickness of metamaterial are deep-subwavelength scaled ($< 0.37\lambda/\text{Re}(n)$). Also, AuNP superlattice was encapsulated by dielectric matrix of epoxy (the real part of refractive index ($n$) of 1.52); the nonlocal effect [6,22] was excluded.

*3.2 Achieving unnaturally high effective permittivity*

First, I explored the effective permittivity and permeability of the colloidal superlattice monolayer, constructed with three AuNP motifs: (i) 50 nm Au nanospheres (NS, $\delta < d$), (ii) 50 nm nanocubes (NC, $\delta < d$), and (ii) nanorectangles (NR) with 50 nm width and 20 nm thickness ($\delta \approx d$). To isolate the effect of AuNP shape on the effective permittivity and permeability, I fixed the lateral dimension of each AuNP to be 50 nm. Also, as AuNPs are inherently capped by 1 nm thick organic ligand [20,23], the minimal gap between AuNPs was set to be 2 nm. The physical thickness of dielectric matrix was also fixed to 80 nm herein. According to the packing rule [19], the lattice geometry of NS monolayer was hexagonal, while that of NC and NR monolayers was tetragonal.

Fig. 2 presents the obtained effective permittivity $\varepsilon = nz_0/z$, where $z_0$ and $z$ are the impedance in free space and in artificial medium, respectively. The AuNP superlattice embedded within dielectric matrix can arise TE waves (i.e., gap plasmon mode), resulting from the capacitive coupling between AuNPs [4,16,17]. Thus, at the resonant frequency of this capacitive coupling, oppositely different charges are accumulated at the faces of each paired NP, resulting in huge dipole moment within the gap of NPs. This large polarization density driven by capacitive coupling between NPs can dramatically increase the effective permittivity. The determinant of this capacitive coupling is the interfacial area between adjacent NPs. As expected, NC with 250 nm$^2$ surface contact is advantageous over both NS with point contact and NR with 100 nm$^2$ surface contact in terms of increasing effective permittivity: the order of effective permittivity at the wavelength of dipolar resonance (i.e., fundamental mode) follows the interfacial area between NPs: ($\varepsilon_{NS}$=29.1) < ($\varepsilon_{NR}$=97.5) < ($\varepsilon_{NC}$=190.7). In the case of NC superlattice, the 2$^{nd}$-order mode of the electric resonance (i.e., quadrupolar resonance) is also observed at higher frequency.

Fig. 3-4 detail such gap plasmonic mode hosted by NP superlattice, when incident electric field is oriented along *x*-axis. The dispersion curve (Fig. 3(a)) confirms that both TE and TM modes can be supported by three NP superlattices; compared to TM counterparts, TE modes are much far from the light cone owing to their strongly trapped features within the gap via capacitive coupling. This is reason that TE parallel momentum of NC superlattice can be higher than those of NS and NR counterparts. In contrast to the order of effective permittivity, NS superlattice shows slightly higher the parallel momentum of TE mode than NR superlattice, mainly due to its different lattice geometry.

Fig. 3(b)-3(d) demonstrate the electric field distribution of TE mode for three NP superlattices. As briefly mentioned above, the incident electric field accumulates the charges at each edge of the faced NPs, in that the huge dipole moment is generated within the gap of NPs. This aspect is well revealed by the strongly confined electric field across the gap; in particular, the insets of Fig. 3(c)-3(d) elucidate the nature of TE mode (i.e., the surface plasmon polaritons-like mode with the vertically oriented electric field with respect to the surface of NP superlattice) [24].

Several important features are further noteworthy. First, the direction of such strongly confined electric field is exactly opposite to that of incident electric field, as an evidence of the capacitive effect. Second, the strength of the confined electric field is much stronger than that of incident light, confirming the huge enhancement of the effective permittivity via electric resonance. As with these vector analyses of electric field, the intensity of electric field ($|E|^2$) is maximized within the gap (see the spatial distribution of the saturated $|E|^2$, shown in Fig. 4). In the case of NC superlattice, the $2^{nd}$-order mode of electric resonance is also observed (Fig. 4(d)). Third, such confined electric field can deeply penetrate into the depth of the gap at the electric resonance wavelength. Especially, the strength of the electric field, which is confined within the gap of NC superlattice ($\delta < d$), is gradually increased along the depth; indicating the propagation of the confined electric field as a form of Fabry-Perot resonance in a longitudinal direction [25,26]. These analyses show how the NP superlattice can harness the extremely high permittivity.

*3.3 Suppressing the diamagnetic response*

Another key to high refractive index is to minimize the diamagnetic effect [16,17]; the geometrical factor including AuNP shape and $\delta/d$ can effectively affect the diamagnetic effect and associated permeability ($\mu=nz/z_0$) (Fig. 5). Fig. 5(a) shows the numerically retrieved real part of permeability for three NP superlattices. A magnetic anti-resonance is obvious ($d\mu/d\lambda$ is opposite to $d\varepsilon/d\lambda$), as another evidence of electric resonance [16,27]: herein, the dip of Re($\mu$) is the indicator of the diamagnetic effect. For the case of NC superlattice, the $2^{nd}$-order of a magnetic anti-resonance is additionally observed at wavelength of 920 nm. Also, the imaginary parts of the effective permeability are opposite in sign, compared with those of effective permittivity, even if not shown in the manuscript for clarity.

More importantly, even single layer of NP superlattice can show the diamagnetic response particularly for $\delta < d$, since the current loop, perpendicular to the incident magnetizing field, can be efficiently harnessed enough. Actually, the NC superlattice ($2\delta \sim d$) leads to a significant diamagnetic response, as confirmed by the lowest dip of Re($\mu$) (~ 0.22 at fundamental mode); this is further confirmed by the suppressed penetration of magnetic field $|H|^2$ (the penetration depth of ~ 24 nm, shown in Fig. 5(d)).

Also, the direct comparison of NC superlattice with NR counterpart allows us to systematically figure out the effect of $\delta/d$ on the diamagnetic effect. The decrease in current loop by simple increase in $\delta/d$ effectively suppresses the diamagnetic effect, so as to increase the dip of Re($\mu$) (~ 0.38 at fundamental mode). Also, this enhanced dip of Re($\mu$) is further confirmed by the increased penetration depth of $|H|^2$ (~ 43 nm).

Another determinant in controlling the diamagnetic effect is the NP shape and lattice geometry. Even if $d$ is same with NC, NS superlattice exhibits much suppressed diamagnetic effect (i.e., the dip of Re($\mu$) is 0.51 at fundamental mode) for following reason. The hexagonally close-packed NS inherently has the four point contacts per each NS in contrast to both NC and NR superlattices. Thus, this structural feature makes the current loop more tortuous; the diamagnetic effect of NS superlattice can be efficiently suppressed.

*3.4 High refractive index of NP superlattice metamaterial*

Fig. 6(a) summarizes the numerically retrieved the real part of refractive index ($n = \sqrt{\varepsilon\mu}$) for three NP superlattices (overall dielectric thickness of 80 nm); as with the results of effective permittivity and permeability, the refractive index is maximized at the wavelength of electric resonance. The obtainable peak refractive indexes of these NP superlattices (i.e., at fundamental mode) are obviously beyond that of dielectric matrix (~ 1.48). The order of peak refractive index at fundamental modes is as follows: (Re($n$)$_{NS}$=4.9) < (Re($n$)$_{NR}$=8.1) < (Re($n$)$_{NC}$=9.2).

Despite of a relative low effective permittivity, the peak refractive index of NC superlattice at $2^{nd}$-order mode is highest (Re($n$)=11.0), due to the negligible diamagnetic effect (Re($\varepsilon$) ≈ 1.0). Also, it is important to note that at the quasi-static regime (satisfying the condition, in which the concentrated electric field within the gap is cancelled by applying electric field), the achievable refractive index still lies in the unnatural regime: (Re($n$)$_{NS}$=3.7) < (Re($n$)$_{NR}$=4.9) < (Re($n$)$_{NC}$=6.8). Thereby, the unnaturally high refractive index at the broadband optical frequencies (100 THz~300 THz) can become available by a versatile assembly of NPs.

Due to its plasmonic resonant behavior, the loss is inherently involved in high-index NP metamaterial. Fig. 6(b) presents the figure of merit (FOM, defined by Re($n$)/Im($n$)) for three NP superlattices. Along the wavelength of interest, FOM is gradually increased after touching the dip around at the electric resonance; eventually, approaching to about 90 (for all cases). Thus, high refractive index together with a low loss can be achieved at the quasi-static limit.

*3.5 Assignment of effective thickness*

So far, I have retrieved the effective optical parameters of NP metamaterial with the 80 nm thick dielectric matrix. However, such refractive index, retrieved with the restriction on 80 nm physical thickness, should be adjusted to satisfy no explicit physical boundaries (referred to as effective thickness estimation) [28]. Particularly, the reflected and transmitted light should be close to a plane wave within the effective thickness of single-layered metamaterial; this condition can be confirmed by the invariant refractive index of bulk medium with respect to the number of the stacked single NP metamaterial layer [17].

Herein, the assignment of effective thickness of the single-layered metamaterial was done only with the NR superlattice (Fig. 7). The overall thickness of multilayered NS and NC superlattices is comparable to the effective wavelength of normally incident light: for example, the minimum thickness of two-layered NS or NC superlattice (~ 100 nm without dielectric matrix) is already about $0.5\lambda/\text{Re}(n)$ at fundamental mode. Thus, the use of homogenization theory and *s*-parameter extraction cannot be justified for NS or NC superlattice.

The bulk refractive index of the NR superlattice, embedded within 30 nm thick dielectrics, is maintained almost unchanged at the quasi-static limit (~ 5 % variation), as the number of layer is increased up to three (see Fig. 7(a)). Beyond four layers, overall artificial medium cannot be analyzed no longer by homogenization theory and *s*-parameter extraction. Thus, I can assign the effective thickness of the single-layered NR superlattice metamaterial to be 30 nm; this deep-subwavelength scaled thickness ($< 0.37\lambda/\text{Re}(n)$) can justify the use of homogenization theory and *s*-parameter extraction. From these results, the effective refractive index of the single-layered NR superlattice metamaterial is found to be 15.7 at the peak and 7.3 at the quasi-static limit, which are far beyond the upper limit of naturally occurring material (e.g., silicon with Re(*n*) of 3.45). The corresponding set of other effective parameters including permittivity, permeability, and FOM is summarized in Fig. 7(b)-7(d).

Then, I addressed the dependence of the effective refractive index on the gap between NRs, which is a pivotal parameter in the capacitive effect. The 2 nm gap is the shortest width due to the geometrical limitation of organic ligand-encapsulated NRs; thus, the fundamental upper limit of the effective refractive index is 15.7 at the peak and 7.3 at the quasi-static regime. As the gap is increased, the capacitive effect gets weaken. Thus, the effective refractive index becomes reduced by increasing the gap (Fig. 7(e)-7(f)). Especially, when the gap is smaller than *d* (i.e., 20 nm), the pair of NRs can behave as the parallel plate capacitor. Thereby, within this parallel plate capacitor regime, the reduction of the gap allows the NR superlattice metamaterial to increase the effective refractive index drastically (~ $(1/\text{gap})^{1/2}$).

*3.6 Electrical control of high refractive index by graphene-NR superlattice metamaterial*

Finally, I explored the possibility for the electrical control of the high refractive index by using graphene-NR superlattice metamaterial. At the wavelength of interest, the complex permittivity of graphene is determined both by interband and by interband transitions. In particular, the real part of graphene permittivity is reversed from positive to negative after touching the peak of the interband transition (Fig. 8(a)). This interband transition and the resultant increase in the real part of graphene permittivity are maximized, when the energy of incident light becomes twice of Fermi-level ($2E_F$). Beyond the peak of interband transition, the real part of graphene permittivity is monotonically decreased and dominantly governed by intraband transition. Meanwhile, the imaginary part of graphene permittivity gets maximized at the intraband transition as well (Fig. 8(b)); beyond this point, intraband transition also acts as a pivotal role in the imaginary part of graphene permittivity. More importantly, this complex permittivity of graphene can be modulated by adjusting $E_F$ of graphene (e.g., gating), in that the incorporation of graphene directly onto the surface of NR superlattice allows electrical tuning of the capacitive effect and the resultant effective refractive index.

First, the attachment of graphene onto the NR superlattice leads to the perturbation of electric resonance between NRs; thus, the capacitive effect gets dimmed spontaneously. In other words, graphene herein acts as an extra dissipation channel. As a result, this graphene hybridization results in the reduction of refractive index (30 nm thick dielectric layer) at electric resonance (from 15.7 to 12.5) (Fig. 8(c)).

Then, I modulated the effective refractive index of this NR superlattice metamaterial by controlling $E_F$ of graphene. Herein, I assumed that the $E_F$ can be reduced until – 700 meV (graphene doping of ~ 9 × $10^{16}$ $cm^{-2}$), which is about the highest doping level in an experimental way (e.g., ion-gel doping) [29]. As summarized in Fig. 8(c)-8(d), such changes of graphene complex permittivity with respect to $E_F$ is well reflected in the modulation of effective refractive index, as follows:

By reducing $E_F$ less than - 200 meV, the real part of graphene permittivity becomes more negative at the wavelength higher than the peak of interband transition (i.e., governed by intraband transition); the charges to be accumulated at the edges of the faced NRs can be leaked to graphene. Thus, the capacitive effect is weakened as the $E_F$ is reduced. Indeed, the decrease in $E_F$ results in the reduced refractive index at the quasi-static limit.

The real part of graphene permittivity can be positively maximized at the wavelength of interband transition. Particularly, when $E_F$ is -600 meV, the peak position of Re($\varepsilon$) of graphene is quite closed to the wavelength of electric resonance of NR superlattice. Thereby, the refractive index at the peak is maximized. Also, the wavelength of electric resonance is slightly shifted toward the position of interband transition. For example, the wavelength of electric resonance is red-shifted, when shorter than the wavelength of interband transition.

**4. Conclusion**

Using AuNP superlattice, the unnaturally high refractive index at broadband optical frequencies can be conceived. Particularly, the use of sub-100-nm NPs allows a versatile access to the metamaterial working in the optical domain. More importantly, the flexibility of controlling NP motifs (i.e., shape and dimension) forces me to dramatically increase the capacitive effect and associated permittivity of the artificial medium, while simultaneously suppressing the diamagnetic response as possible. Furthermore, the assembly of NPs could be highly compatible with the graphene transfer; thereby, the high refractive index at broadband optical frequencies can be tunable through electrical gating.

The strategy demonstrated in this work is useful as modulating broadband optical wave and also expanding transformation optics at optical frequencies.

**Acknowledgements**

This work was fully supported by Samsung Research Funding Center for Samsung Electronics under Project Number SRFC-MA1402-09.

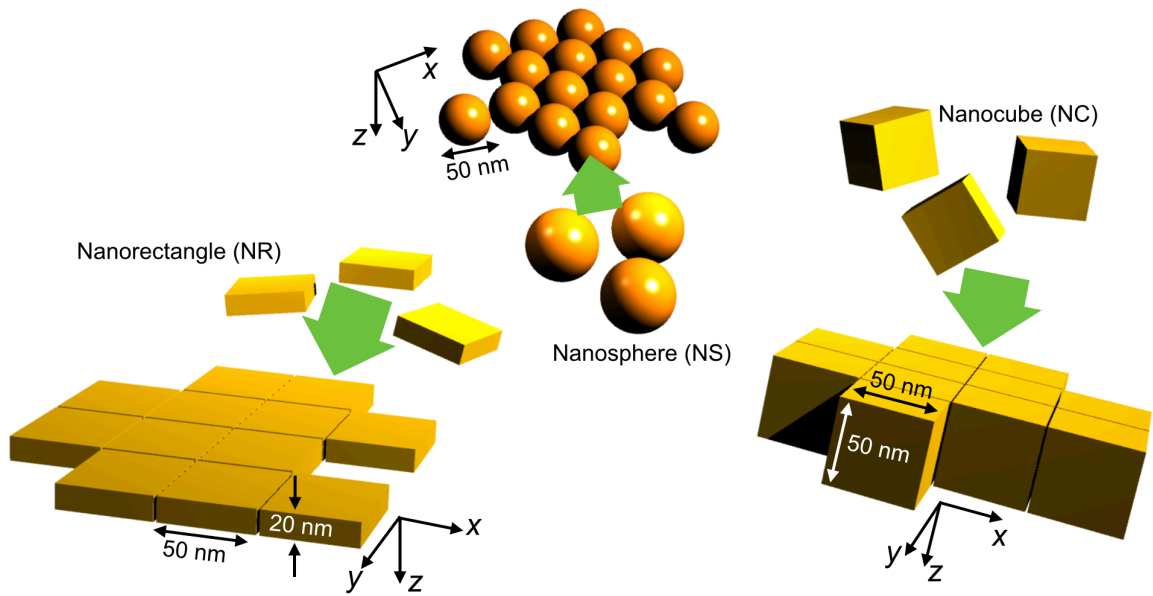

**Figure 1.** Schematic for the assembly of gold nanoparticles (AuNPs) into superlattice metamaterials with nanosphere (NS), nanocube (NC), and nanorectangle (NR).

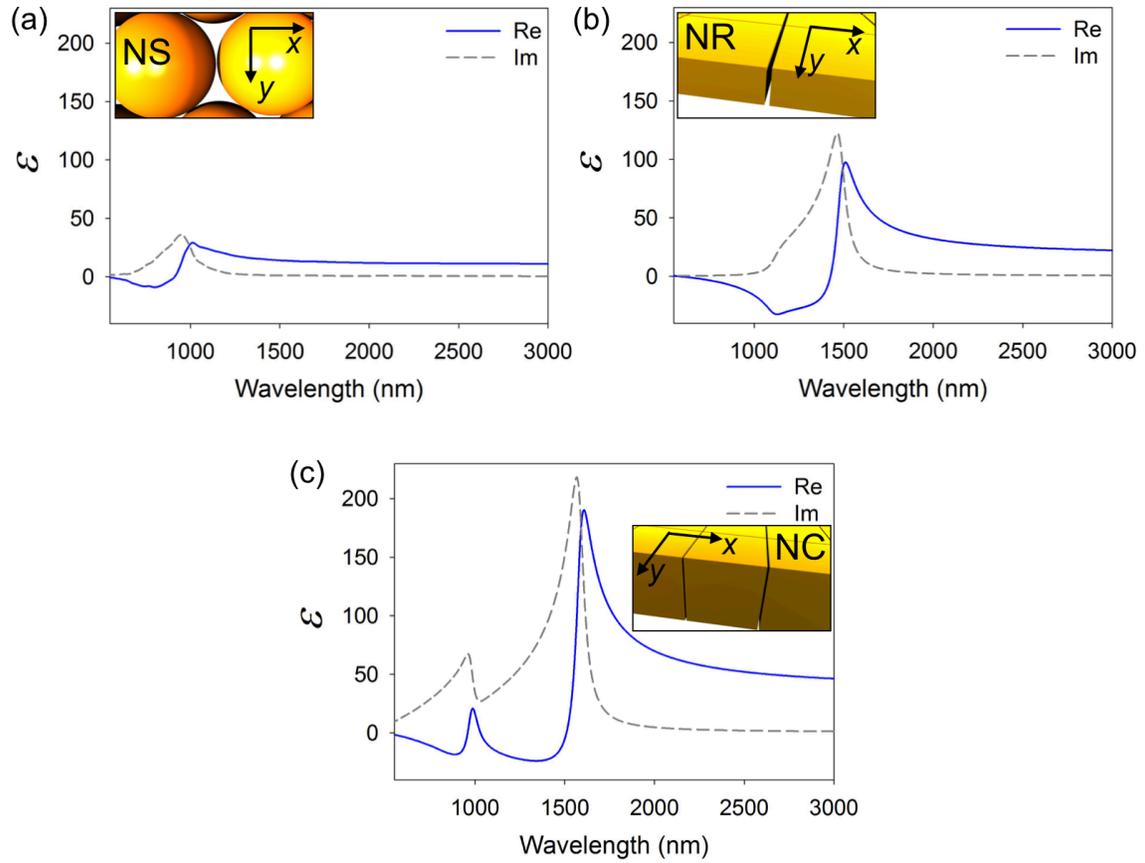

**Figure 2.** Numerically calculated effective permittivity of NP superlattice metamaterial (overall thickness of dielectric matrix is 80 nm): (a) NS, (b) NR, and (c) NC superlattices.

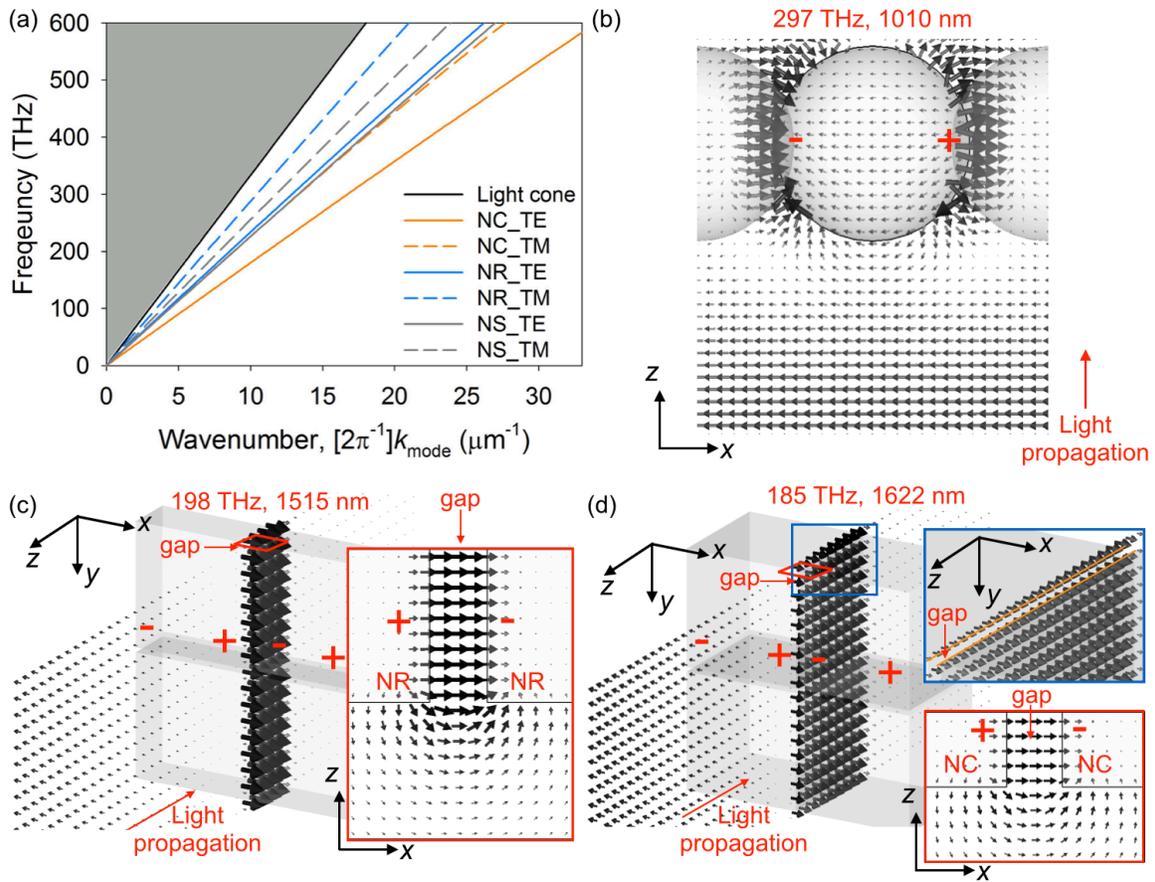

**Figure 3.** (a) Dispersion curves of NP superlattice metamaterial, constructed by assembly of NS, NR, and NC motifs (overall thickness of dielectric matrix is 80 nm). (b-d) The spatial distributions of electric field in (b) NS, (c) NR, and (d) NC superlattices at electric resonance frequencies (i.e., fundamental mode).

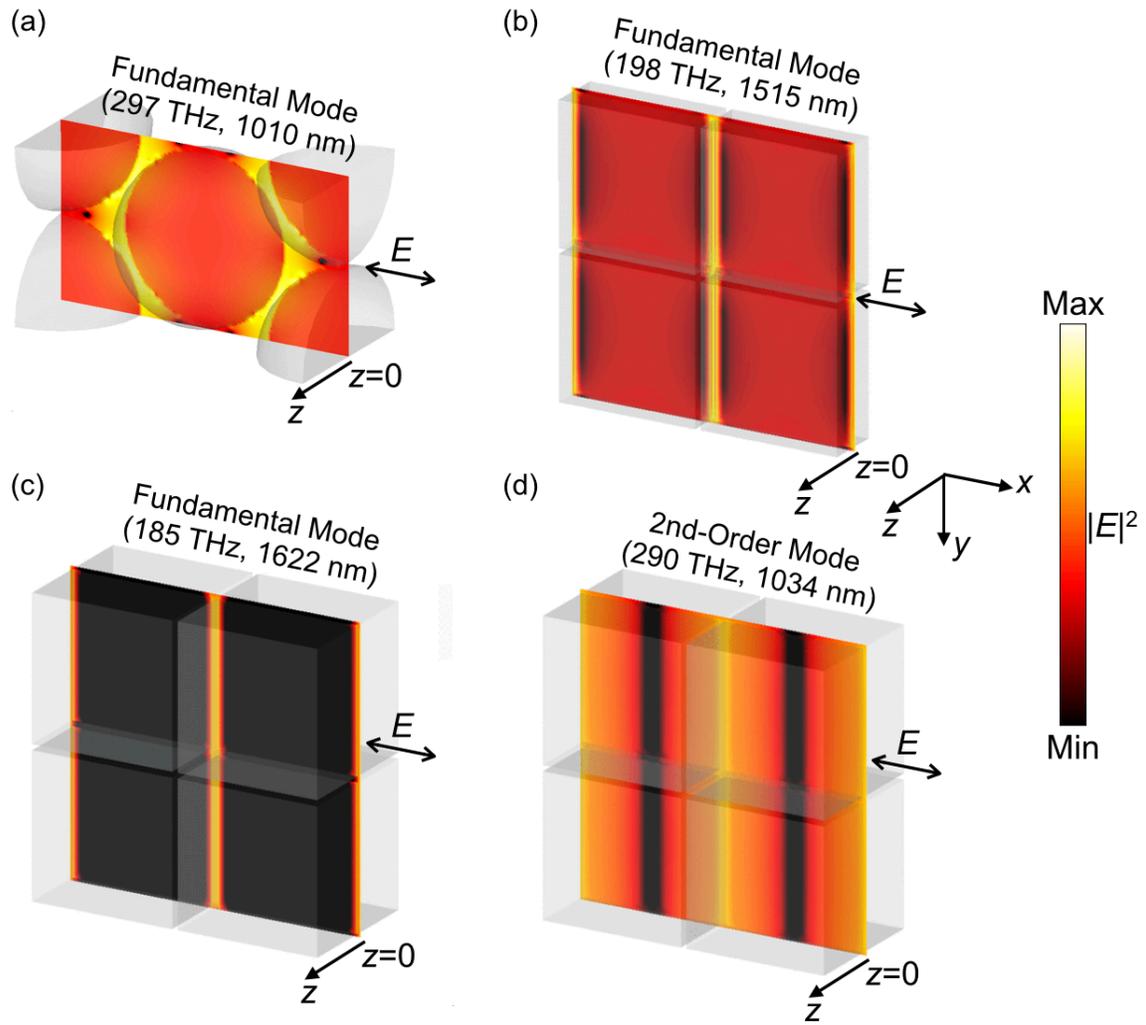

**Figure 4.** Spatial distribution of the saturated electric field intensity ($|E|^2$): (a) for NS superlattice at fundamental electric resonance frequency, (b) for NR superlattice at fundamental electric resonance frequency, and (c-d) for NC superlattice at (c) fundamental and (d) 2nd-order mode frequencies.

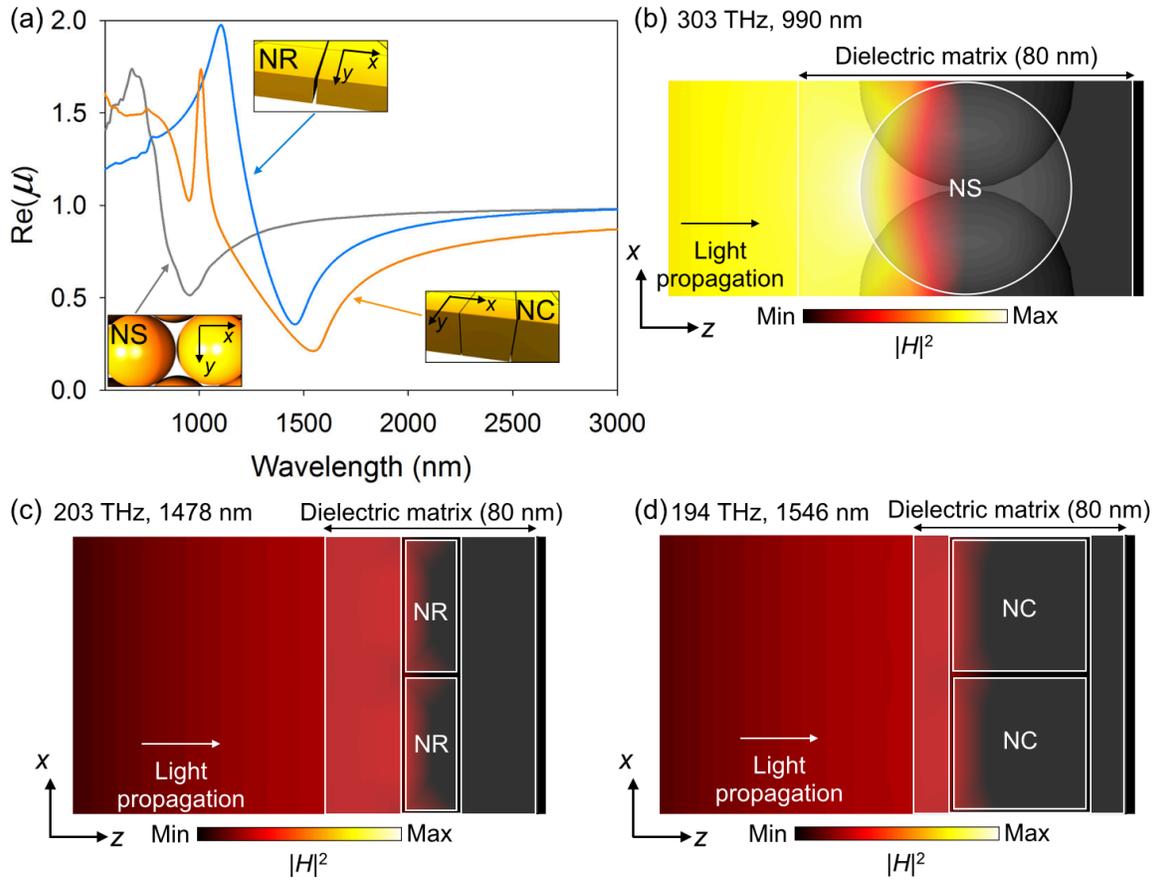

**Figure 5.** (a) Numerically calculated effective permeability of NS, NR, and NC superlattices (overall thickness of dielectric matrix is 80 nm). (b-d) Spatial distribution of saturated magnetic field intensity ($|H|^2$) for (b) NS, (c) NR, and (d) NC superlattices at fundamental anti-magnetic resonance frequencies.

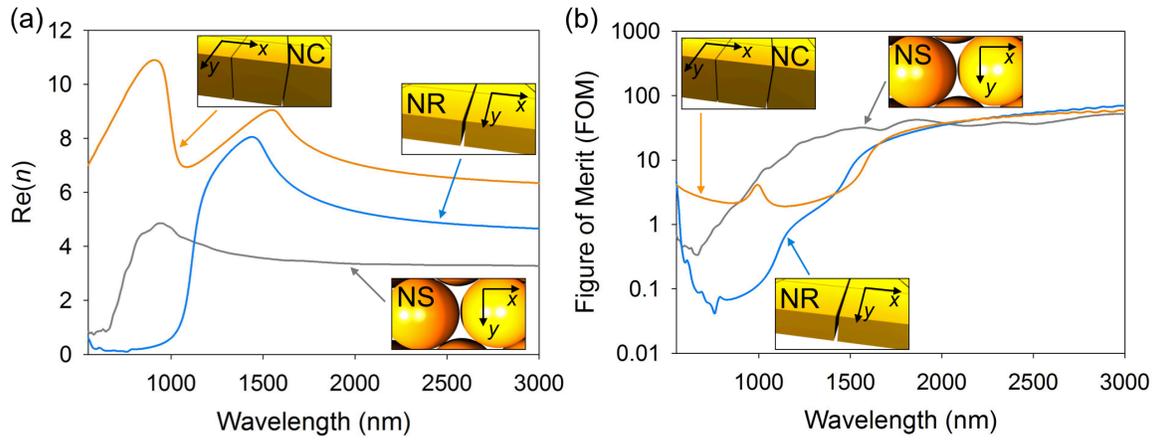

**Figure 6.** Numerically calculated (a) effective refractive index and (b) figure of merit (FOM defined by Re(*n*)/Im(*n*)) for NS, NR, and NC superlattices (overall thickness of dielectric matrix is 80 nm).

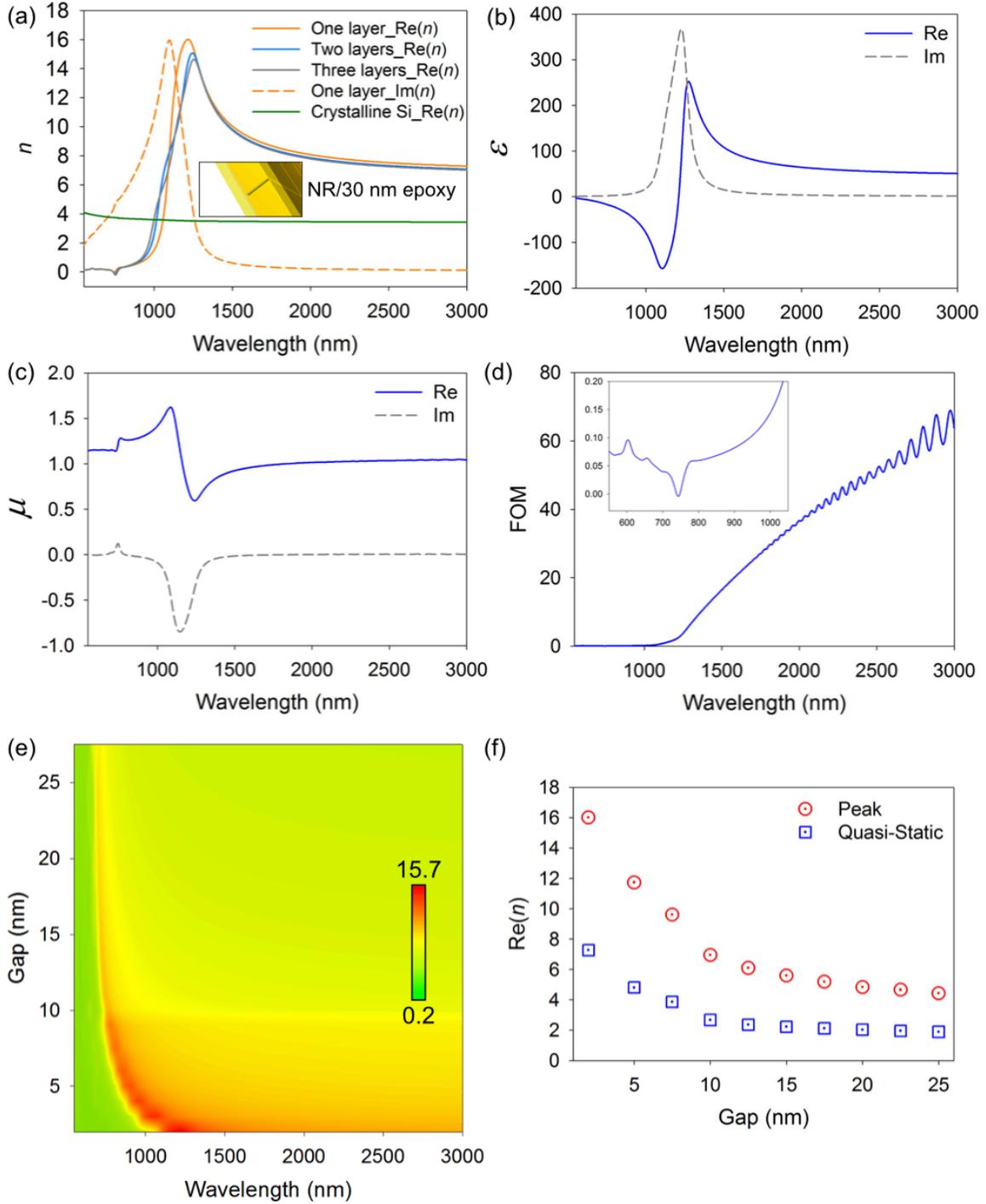

**Figure 7.** Numerically calculated (a) effective refractive index, (b) permittivity, (c) permeability, and (d) FOM for NR superlattice metamaterial with the assigned effective thickness of 30 nm. (e-f) The dependence of the effective refractive index on the gap between the paired NRs.

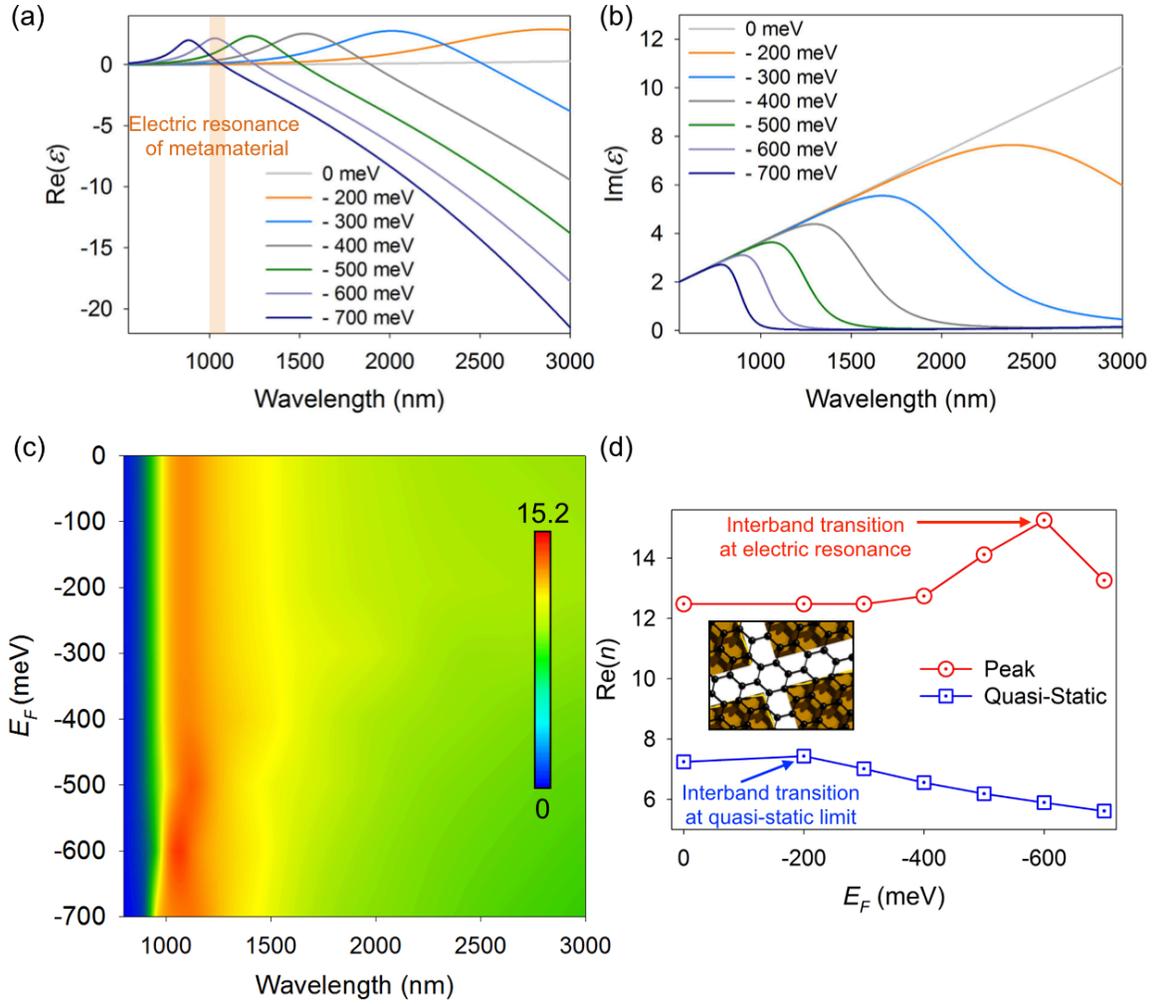

**Figure 8.** Electrical modulation of the complex permittivity of graphene with respect to the Fermi-level ($E_F$): (a) real part and (b) imaginary part. Modulation of the refractive index of NR-superlattice metamaterial with respect to $E_F$: (c) refractive index contour map and (d) value of refractive index at the peak and at the quasi-static limit (i.e., wavelength of 3000 nm).